# COSMOLOGICAL MODEL WITH A LOCAL VOID: NEW SUPERNOVA CONSTRAINTS

LE TUAN ANH HO

*Department of Physics, National University of Singapore, 2 Science Drive 3*
*Singapore 117542*

SHAO CHIN CINDY NG

*Department of Physics, National University of Singapore, 2 Science Drive 3*
*Singapore 117542*

A simple inhomogeneous cosmological model with a local void is constrained with the latest Union supernova compilation. To fit the supernova data, a large local void on the scales of 1 Gpc is found, contrary to the small scales of 200 Mpc in the previous finding. A more realistic inhomogeneous cosmological model may be required to fit the supernova data. Alternatively, a clumpy universe with $\alpha < 1$ can fit the supernova data with reduced local void scales.

## 1 Introduction

Nowadays, with tremendous development of technology, more and more precise cosmological observations are being performed. Type Ia supernovae (Sne Ia), from which we can derive the magnitude-redshift ($m$-$z$) relation, are among the observations that play important roles in constraining the cosmological parameters. Analyzed in the Friedmann framework, a model of the Universe with nearly 70% dark energy and 30% matter [1-3] has risen. This model is popularly called the concordance model.

The concordance model is based on the homogeneity and isotropy of space. This model has obtained many impressive successes. Apart from the Sne Ia, it can well explain and predict many other observations, including the cosmic microwave background (CMB) anisotropy and baryon acoustic oscillation (BAO). However, the model also encounters several unresolved issues, such as the coincidence problem, cosmological constant problem, and physical nature of the dark energy.

Recently, evidences from the number count of galaxies [4] and an earlier data analysis of Sne Ia [5] have shown that we may live in a local void. Furthermore, many voids with different sizes, and several huge nonlinear structures have also been revealed through surveys like SDSS and 2dFGRS [6, 7]. Voids have also been used to explain cold spots [8] and some features of low multipole anomalies [9] in the CMB data. On those bases, several authors have proposed cosmological models with a local void to interpret cosmological observations without the dark energy component [10-17] (for a review, see [18]). These authors dropped the Copernican principle and supposed that we live near the centre of a local void. Most of them used the LTB model as toy models for data fitting.

In 1990s, Tomita proposed a simple model with a local void on scales of about 200 Mpc [11, 19, 20]. The author showed that the HSST [1, 2] and the SCP data [3] could be fitted in this framework without a cosmological constant. Since then many years have past, and it is necessary to re-examine Tomita's model with the latest datasets. In addition, Tomita's models with clumpiness parameter [21, 22] different from unity need to be further investigated.

In this paper, we constrain the matter and cosmological constant densities in Tomita's model using the Union supernova compilation (see Kowalski et. al. [23]). In §2, we briefly review the Tomita's model and Sne Ia data-fitting procedure. In §3, we constrain the matter and cosmological constant densities. In §4, we constrain the models with clumpiness parameter smaller than unity. And finally, a conclusion and discussion are given in §5.

## 2  Tomita's Model and Sne Ia Observations

As mentioned in the introduction, we may live in a local void and are surrounded by huge nonlinear structures. Moreover, some evidences show that there may exist inhomogeneities in the global Hubble flow [24, 25]. On those bases, Tomita [11, 19, 20] proposed a simple spherically symmetric and inhomogeneous cosmological model. This model consists of two regions: inner and outer. The inner matter density ($\Omega_{in}$) is smaller than the outer one ($\Omega_{out}$), and the two regions are separated by a spherical singular shell whose mass compensates the inner region mass deficiency. Correspondingly, the inner Hubble constant ($H_{in}$) is larger than the outer one ($H_{out}$), which provides a different viewpoint to accelerating expansion in the concordance model with dark energy. In Tomita's model, accelerating expansion is only a consequence of expansion rate variation at the distance near the observer, and overall the Universe is decelerating. In another words, accelerating expansion is only an apparent phenomenon.

In this model, the observer is assumed to be at the centre of the local void. The existence of a centre seems to be non-philosophical, but it is necessary to note that the centre belongs to the local void, not the whole Universe. The outer region is assumed to be homogeneous to avoid over-complication. Beyond the local void, there actually lie many other voids in the universe. A homogeneous outer region is a simplified but reasonable assumption, because when we interpret observations, the local void affects the cosmological measurements the most.

Here, we follow Tomita's formalism [10, 11] for fitting the theoretical model with the Sne Ia data. The differential equations and corresponding boundary conditions for calculating angular diameter distance have been given in Eqs. (5) – (11) of Ref. [11]. The luminosity distance ($D_L$) is readily calculated from the angular diameter distance ($D_A$) and redshift ($z$): $D_L = (1 + z)^2 D_A$. The distance modulus $\mu = 5 \log (D_L/\text{Mpc}) + 25$. Utilizing $\chi^2$ statistic, we are able to constrain the parameters of the model. The details of the method have been described in [10].

For Sne Ia dataset, we will use the Union compilation (Table 11 in [23]), which is updated and uses only one analysis procedure in the whole compilation. Following Kowalski *et al.* [23], throughout this work, we only use the 307 Sne Ia that pass 3σ outlier cut.

## 3 Standard Parameters and Confidence Contours

Tomita's model has 7 parameters, Hubble constants and matter densities in the inner and outer regions, $H_{in}$, $H_{out}$, $\Omega_{in}$, $\Omega_{out}$, outer dark energy density in the form of cosmological constant, $\Omega_\lambda$, the redshift of the void boundary, $z1$, and the clumpiness parameter, $\alpha$. In order to avoid over-complication, we will follow [10, 11] and examine $\Omega_{out}$ and $\Omega_\lambda$ for some specific values of $R \equiv H_{out} / H_{in}$, $z1$, $\Omega_{in}$, and $\alpha$.

First, we consider matter density profile A for $\Omega_{in}$ (see Table 1) and $\alpha = 1$. For $\Omega_\lambda = 0$, we find the following standard parameters' values: $R = 0.69$ and $z1 = 0.23$. The confidence contours for the standard model with the above mentioned parameters are shown in Figure 1.

Table 1. Four Matter Density Profiles

| Profile | $\Omega_{in}$ |
|---|---|
| A | $\Omega_{in} = \Omega_{out} / 2$ if $\Omega_{out} < 0.6$; $\Omega_{in} = 0.3$ if $\Omega_{out} \geq 0.6$ |
| B | $\Omega_{in} = \Omega_{out} R^2$ |
| C | $\Omega_{in} = 0.3$ for all $\Omega_{out}$ |
| D | $\Omega_{in} = 0.2$ for all $\Omega_{out}$ |

For comparison, we also obtain the confidence contours for the standard parameters $(R, z1) = (0.80, 0.08)$ in Tomita's earlier analysis [10]. The confidence contours with the new Union supernova compilation are presented in Figure 2. We find $(\Omega_{out}, \Omega_\lambda) = (1, 0)$ lies outside the 2σ-confidence contour. The density parameters $(\Omega_{out}, \Omega_\lambda) = (1, 0)$ correspond to a flat outer region without a cosmological constant.

In Figures 3 and 4, we present the confidence contours when the parameters are varied from their standard values. Specifically, we present the confidence contours for $R = (0.65, 0.69, 0.73)$ and $z1 = (0.21, 0.23, 0.25)$. In general we find the contours vary, in a way similar to the finding in Tomita's earlier analysis [10]. As shown in Figure 3, as $R$ increases the confidence contours move in the direction of decreasing $\Omega_{out}$ and increasing $\Omega_\lambda$. This variation can be readily explained: as $R$ (the Hubble contrast) approaches 1, the level of inhomogeneity of the model decreases, and the confidence contours approach those of the homogeneous concordance model [23]. In Figure 4, we find the confidence contours moving in the direction of increasing $\Omega_{out}$ and $\Omega_\lambda$ as $z1$ increases. When the local void size increases, the spatial curvature for the outer region also increases accordingly.

In Table 2, we list the best-fit $\Omega_{out}$, $\Omega_\lambda$ and minimum $\chi^2$ for the confidence contours in Figures 3 and 4. The $\chi^2$ per degree of freedom (dof = 305) only varies a little from one model to another.

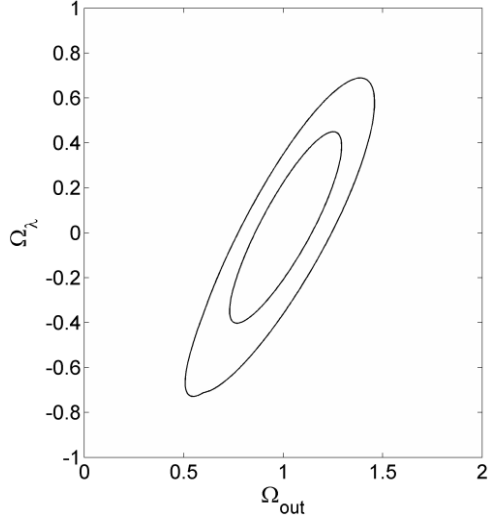

Figure 1. 68.3% and 95.4% confidence contours in $\Omega_{out} - \Omega_\lambda$ plane, for $R = 0.69$, $z1 = 0.23$, matter density profile A, and $\alpha = 1$.

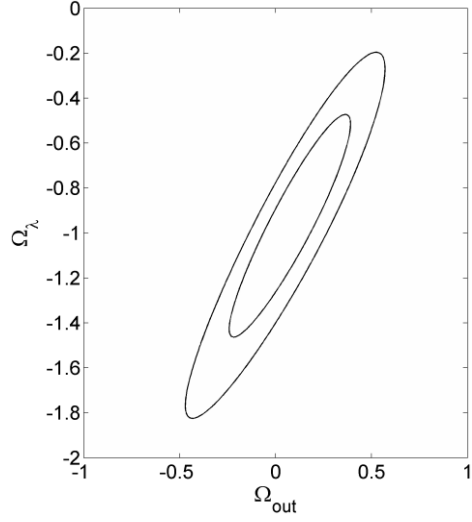

Figure 2. 68.3% and 95.4% confidence contours in $\Omega_{out} - \Omega_\lambda$ plane, for $R = 0.80$, $z1 = 0.08$, matter density profile A, and $\alpha = 1$. Note that $(\Omega_{out}, \Omega_\lambda) = (1, 0)$ lies outside the 2σ-confidence contour.

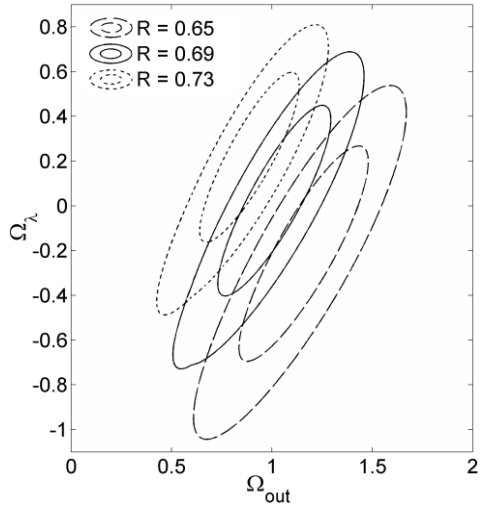

Figure 3. 68.3% and 95.4% confidence contours in $\Omega_{out} - \Omega_\lambda$ plane, for $R = (0.65, 0.69, 0.73)$, $z1 = 0.23$, matter density profile A, and $\alpha = 1$.

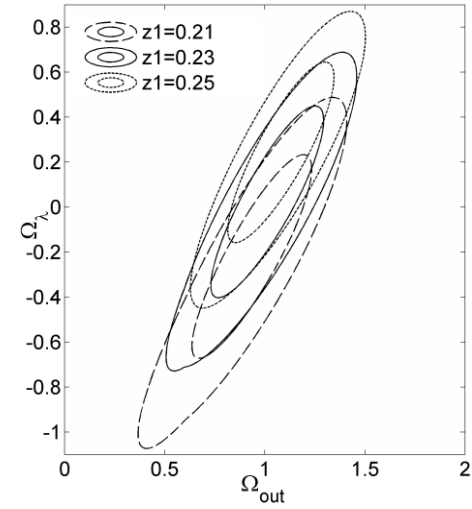

Figure 4. 68.3% and 95.4% confidence contours in $\Omega_{out} - \Omega_\lambda$ plane, for $R = 0.69$, $z1 = (0.21, 0.23, 0.25)$, matter density profile A, and $\alpha = 1$.

Table 2. Best-fit $\Omega_{out}$, $\Omega_\lambda$, and minimum $\chi^2$ for different values of $R$ and $z1$.

| $R$ | $z1$ | $\Omega_{out}$ | $\Omega_\lambda$ | $\chi^2_{min}$ |
|---|---|---|---|---|
| 0.65 | 0.23 | 1.17 | -0.19 | 317.51 |
| 0.69 | 0.21 | 0.95 | -0.19 | 316.54 |
| 0.69 | 0.23 | 1.02 | 0.04 | 316.04 |
| 0.69 | 0.25 | 1.09 | 0.26 | 315.42 |
| 0.73 | 0.23 | 0.90 | 0.24 | 314.66 |

In this work we also constrain $\Omega_{out}$ and $\Omega_\lambda$ for the different density profiles listed in Table 1. The results are summarized in Table 3. Interestingly, profile C (with constant $\Omega_{in}$ = 0.3) has the same best-fit values as profile A. Profile B (with different inner and outer Hubble constants but equal inner and outer matter densities, $\rho_{in}$ and $\rho_{out}$) and profile D (with constant $\Omega_{in}$ = 0.2) give best-fit values only slightly different from those of profile A. We thus assume that the density profile has little effect on constraining $\Omega_{out}$ and $\Omega_\lambda$.

Table 3. Best-fit $\Omega_{out}$, $\Omega_\lambda$, and minimum $\chi^2$ for different matter density profiles, for $R = 0.69$ and $z1 = 0.23$.

| Profile | $\Omega_{out}$ | $\Omega_\lambda$ | $\chi^2_{min}$ |
|---|---|---|---|
| A | 1.02 | 0.04 | 316.04 |
| B | 1.01 | 0.06 | 317.28 |
| C | 1.02 | 0.04 | 316.04 |
| D | 1.00 | 0.00 | 315.49 |

## 4  Clumpy Universe and Local Void Size

The clumpiness parameter is a parameter in Dyer-Roeder's distance equation for clumpy universe [21, 22]. In Tomita's earlier analyses of Sne Ia [11, 20], the author assumed the Friedmann distance ($\alpha = 1$). In this work, we also consider the general distances when $\alpha \neq 1$.

The 68.3% confidence contours for $\alpha = (0.25, 0.5, 0.75, 1)$ are presented in Figure 5. In general, the contour becomes wider as $\alpha$ decreases. Similar to Figure 4, the contour also moves in the direction of increasing $\Omega_{out}$ and $\Omega_\lambda$ as $\alpha$ decreases. For $\alpha = 0.5$, we find that a model with $R = 0.77$ and a small $z1 = 0.16$ provides a reasonably good fit to the Sne Ia data for $(\Omega_{out}, \Omega_\lambda) \approx (1, 0)$ (see Table 4).

In this work, we also consider the more general cases when the inner clumpiness parameter, $\alpha_{in}$, is different from the outer one, $\alpha_{out}$. In Figure 6, we present the confidence contours for $\alpha_{out} = 1$ and $\alpha_{in} = (0.25, 0.5, 0.75, 1)$. We find that the contours almost overlap. As shown in Table 4, the best-fit values only vary a little as $\alpha_{in}$ varies. We thus assume that varying $\alpha_{in}$ from $\alpha_{out}$ has little effect on the constraint.

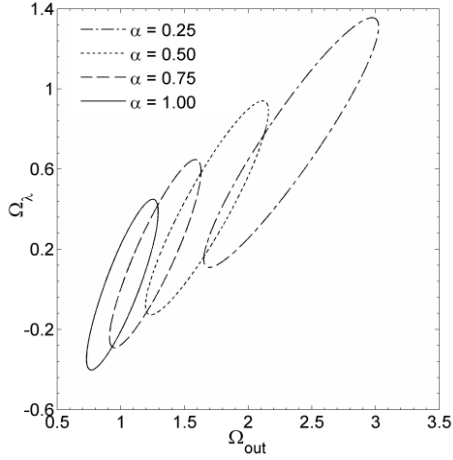
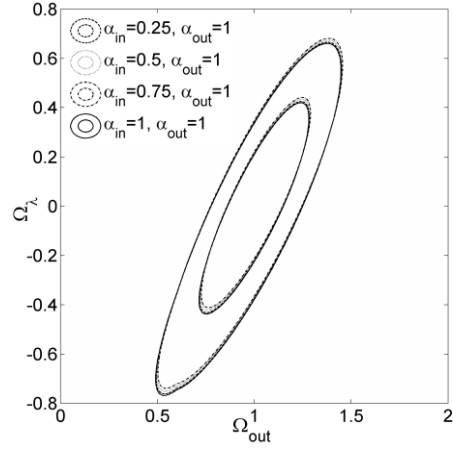

Figure 5. 68.3% confidence contours in $\Omega_{out} - \Omega_\lambda$ plane, for $R = 0.69$, $z1 = 0.23$, matter density profile A, and $\alpha = (0.25, 0.5, 0.75, 1)$.

Figure 6. 68.3% and 95.4% confidence contours in $\Omega_{out} - \Omega_\lambda$ plane, for $R = 0.69$, $z1 = 0.23$, matter density profile A, $\alpha_{out} = 1$, and $\alpha_{in} = (0.25, 0.5, 0.75, 1)$.

Table 4. Best-fit $\Omega_{out}$, $\Omega_\lambda$, and minimum $\chi^2$ for the models in Figures 5 and 6, and a model with $R = 0.77$, $z1=0.16$, and $\alpha_{in} = \alpha_{out} = 0.5$ (the last model).

| $\alpha_{in}$ | $\alpha_{out}$ | $\Omega_{out}$ | $\Omega_\lambda$ | $\chi^2_{min}$ |
|---|---|---|---|---|
| 0.25 | 0.25 | 2.34 | 0.75 | 313.26 |
| 0.50 | 0.50 | 1.68 | 0.42 | 314.25 |
| 0.75 | 0.75 | 1.28 | 0.20 | 315.25 |
| 1.00 | 1.00 | 1.02 | 0.04 | 316.04 |
| 0.25 | 1.00 | 1.01 | 0.02 | 315.88 |
| 0.5 | 1.00 | 1.01 | 0.02 | 315.93 |
| 0.75 | 1.00 | 1.01 | 0.04 | 315.99 |
| 0.50 | 0.50 | 1.01 | 0.06 | 314.66 |

## 5  Conclusion and Discussion

In this paper, we updated the constraint on Tomita's cosmological model with a local void, using the new Union supernova compilation. We found that the Tomita's model with a local void on scales of about 200 Mpc does not provide a good fit to the Sne Ia data when $(\Omega_{out}, \Omega_\lambda) = (1, 0)$. To fit the Sne Ia data when $(\Omega_{out}, \Omega_\lambda) = (1, 0)$, we found a model with $R = 0.69$ and $z1 = 0.23$, corresponding to a local void on scales of about 1 Gpc. Previously, Garcia-Bellido and Haugboelle [14] also fitted the Sne Ia data to the LTB models and found large local voids on the scales of 2.5 Gpc. Recently, Hunt and Sarkar [26] argued that the existence of a large local void in the universe is physically unlikely. However, inhomogeneous cosmological models without dark energy cannot be dismissed on that basis, as the Tomita's model and LTB models are only toy models. A more realistic cosmological model [27-30] may better demonstrate that the accelerating expansion of the universe is only an apparent phenomenon caused by inhomogeneous distribution of matter in space.

In this paper, we also considered the more general case when the distance in Tomita's model is the Dyer-Roeder distance for clumpy universe. We found that for $\alpha$ significantly smaller than 1, a model with local void on smaller scales can provide a reasonably good fit to the Sne Ia data. It remained to be verified if $\alpha$ is indeed smaller than 1 which, when verified, may provide an evidence for a simple inhomogeneous cosmological model with a small local void to explain the Sne Ia observations without the dark energy.

**Acknowledgments**

We would like to thank K. Tomita for helpful discussions.